\documentclass[twocolumn]{revtex4-1}
\usepackage[utf8]{inputenc}
\usepackage[english]{babel}
\usepackage[T2A]{fontenc}
\usepackage{graphicx}
\usepackage{fullpage}
\usepackage{amsmath}
\usepackage{amssymb}
\usepackage{physics}
\usepackage{siunitx}
\usepackage{hyperref}
\usepackage{microtype}
\hypersetup{
    linkcolor=blue,
    citecolor=blue,
    urlcolor=blue,
    colorlinks=true,
    pdfauthor={A. A. Golovanov, E. N. Nerush, I. Yu. Kostyukov},
    pdftitle={Radiation reaction-dominated regime of wakefield acceleration}
}

\newcommand{\avg}[1]{\left\langle #1 \right\rangle}

\newcommand{\plasm}{{\textup{p}}}

\newcommand{\phase}{{\textup{ph}}}
\newcommand{\lorentz}{{\textup{L}}}
\newcommand{\rr}{{\textup{RR}}}

\newcommand{\submax}{{\textup{max}}}
\newcommand{\submin}{{\textup{min}}}

\newcommand{\threshold}{{\textup{th}}}

\newcommand{\schwinger}{{\textup{S}}}
\newcommand{\accelerating}{{\textup{acc}}}
\newcommand{\effective}{{\textup{eff}}}
\newcommand{\relative}{{\textup{rel}}}
\newcommand{\bunch}{{\textup{b}}}
\DeclareMathOperator{\GammaFunc}{Gamma}

\begin{document}

\title{Radiation reaction--dominated regime of wakefield acceleration}
\author{A.\,A. Golovanov}
\author{E.\,N. Nerush}
\author{I.\,Yu. Kostyukov}
\affiliation{Institute of Applied Physics RAS, 603950 Nizhny Novgorod, Russia}

\begin{abstract}
    We study electron acceleration in a plasma wakefield under the influence of the radiation-reaction force caused by the transverse betatron oscillations of the electron in the wakefield.
    Both the classical and the strong quantum-electrodynamic (QED) limits of the radiation reaction are considered.
    For the constant accelerating force, we show that the amplitude of the oscillations of the QED parameter $\chi$ in the radiation-dominated regime reaches the equilibrium value determined only by the magnitude of the accelerating field, while the averaged over betatron oscillations radiation reaction force saturates at the value smaller than the accelerating force and thus is incapable of preventing infinite acceleration.
    We find the parameters of the electron bunch and the plasma accelerator for which reaching such a regime is possible.
    We also study effects of the dephasing and the corresponding change of accelerating force over the course of acceleration and conclude that the radiation-dominated regime is realized both in cases of single-stage acceleration with slow dephasing (usually corresponding to bunch-driven plasma accelerators) and multi-stage acceleration with fast dephasing (corresponding to the use of laser-driven accelerators).
\end{abstract}

\maketitle

\section*{Introduction}

Particle accelerators are now a main tool in laboratory high-energy physics and fundamental particle physics.
To reach new high-energy frontiers, bigger accelerator systems are needed, and their construction is a great technological and financial challenge.
That is why much attention is now attracted to alternative acceleration methods providing extremely high acceleration gradients that may strongly reduce the accelerator cost.
Impressive results are demonstrated by plasma-based methods.
The energy of the accelerated electrons exceeded 8 GeV for a laser driver propagating in a 20-cm-long plasma \cite{Gonsalves_2019_PRL_122_84801}, while the energy doubling up to \SI{85}{\GeV} over the length of \SI{85}{\cm} was observed in beam--plasma experiments \cite{Blumenfeld_2007_Nature_445_741}. 

The physics of plasma-based methods is as follows.
When the driver (an intense laser pulse or a dense charged particle bunch) propagates in a plasma, it excites oscillations of plasma electrons behind it (a plasma wakefield).
If the driver is strong enough (a very intense laser pulse or a bunch with very high charge density), it can cause almost complete evacuation of plasma electrons in some regions behind it \cite{Pukhov_2002_APB_74_355}.
At the same time, heavier ions remain almost immobile.
The perturbation of the plasma electron density leads to violation of quasi-neutrality and to generation of strong accelerating electric fields.
The strength of the accelerating plasma fields can be several orders of magnitude higher than in conventional accelerators \cite{Esarey_2009_RMP_81_1229}. 

Strong fields in accelerating structures may cause significant radiative losses.
In a plasma wakefield, in addition to the longitudinal accelerating force, the transverse focusing force acts on the electron, causing transverse betatron oscillations of the particle.
An electron moving along a curvilinear trajectory emits electromagnetic waves, which leads to radiative losses, thereby reducing acceleration effectiveness \cite{Michel_2006_PRE_74_26501, Kostyukov_2006_JETP_103_800, Kostyukov_2012_PRSTAB_15_111001}.
With typical parameters available in modern plasma acceleration experiments, the radiative losses are usually negligible.
However, with the increase in the electron energy, the radiation reaction effect can be enhanced and affect the operation of plasma-based lepton colliders \cite{Schroeder_2010_PRSTAB_13_101301, Nakajima_2011_PRSTAB_14_91301, Pugacheva_2018_QE_48_291}.
It was shown that the radiative losses can cause the beam emittance self-cooling \cite{Deng_2012_PRSTAB_15_81303}. 

The focusing force structure in plasma wakefield causing the electron trajectory bending is similar to that in beam--beam interaction \cite{Samsonov_2021_arXiv_2107.04787}.
The bending of the particle trajectory leads to synchrotron radiation or beamstrahlung \cite{Blankenbecler_1987_PRD_36_277, Chen_1988_beamstrahlung}.
The total power of the photon emission is governed by the dynamical QED parameter $\chi$ \cite{Ritus_1985_JSLR_6_497, Berestetskii_QE_Vol4, Baier_1989_NPB_328_387}
\begin{align}
    &P (\chi) = \frac{\alpha m^{2}c^{4}}{3 \sqrt{3}\pi\hbar}
    \int_{0}^{\infty}\dd{u} \Lambda (u) K_{2/3}\left(\frac{2u}
    {3\chi}\right), \label{eq:radiationReactionPower} \\
    &\Lambda (u) = \frac{4u^{3}+5u^2+4u}{(1+u)^{4}}, \\
    &\chi = \frac{\gamma}{E_\schwinger}
    \sqrt{\left(\vb{E}+\frac{\vb{v}}{c}\cp\vb{B}\right)^{2}-\left(\frac{\vb{v}}{c}\cdot\vb{\vb{E}}\right)^{2}},\label{eq:chi}
\end{align}
where $K_{\nu}(x)$ is the modified Bessel function of the second kind \cite{Abramowitz_1964_Handbook}, $\vb{v}$ is the particle velocity, $\vb{E}$ and $\vb{B}$  are the electric and the magnetic fields, $E_\schwinger=m^2 c^3 / \hbar e$ is the critical Sauter--Schwinger field~\cite{Berestetskii_QE_Vol4}, $\alpha=e^2/\hbar c$ is the fine-structure constant, $c$ is the speed of light, $\hbar$ is the reduced Planck constant, $m$ and $e>0$ are the electron mass and the elementary charge, respectively.
In both the classical ($\chi \ll 1$) and the strong QED ($\chi \gg 1$) limits Eq.~\eqref{eq:radiationReactionPower} can be reduced to simple power-law expressions
\begin{align}
    &P_{\chi\ll1} (\chi) = \frac{2}{3} \frac{\alpha m^{2}c^{4}} {\hbar}\chi^2,     \label{eq:radiationReactionPowerClassical} \\
    \label{eq:radiationReactionPowerQED}
    &P_{\chi\gg1} (\chi) = C_1 \frac{\alpha m^{2}c^{4}}{\hbar} \chi^{2/3},
\end{align}
where $C_1 = \GammaFunc(2/3)\, 3^{2/3} (2/3)^5 \approx 0.37$, and $\GammaFunc(x)$ is the gamma function \cite{Abramowitz_1964_Handbook}.

In the radiation-dominated regime, the radiation reaction force plays a key role in dynamics of charged particles in extremely strong electromagnetic (EM) field.
For example, in the strong rotating electric field, the particle trajectory is attracted to the curve where the radiative losses are exactly compensated by the work of the electric field \cite{Zeldovich_1975_SPU_18_79, Esirkepov_2015_PLA_379_2044, Kostyukov_2016_PoP_23_93119}. 
For a wide range of parameters and EM field configurations, particles tend to the trajectory providing minimal radiative losses \cite{Gonoskov_2018_PoP_25_93109, Samsonov_2018_PRA_98_53858}.
This asymptotic trajectory corresponds to the self-similar solution of the particle's equation of motion when there is an additional relation between particle parameters, like the particle energy and the absolute value of the radiation reaction force.
For plasma-based acceleration with a linear focusing force and a constant accelerating field (which corresponds the electron staying in the same phase of the wakefield), it was shown that the acceleration of an electron, regardless of the initial conditions, goes into a radiation-dominated (or asymptotic) regime where the averaged radiation reaction force reaches the equilibrium value equal to $2/3$ of the accelerating force \cite{Kostyukov_2012_PRSTAB_15_111001}.
Thus, electron acceleration in the limit $t \rightarrow \infty$ continues indefinitely, but the acceleration rate decreases by a factor of 3 compared to the case of no radiation reaction. 

In real conditions, the accelerating force in a plasma wakefield is inhomogeneous, therefore the force acting on the electron changes as the electron ``slides'' along the wave.
In this paper, we analyze the radiation-dominated regime of particle acceleration taking into account the particle slippage with respect to the wakefield phase.
As a rule, the radiation-dominated regime is realized at extremely high particle energy when the difference between its velocity and the velocity of light can be neglected.
In this case, the slippage of the electron relative to the wake wave will be determined only by the phase velocity of the wakefield.
For laser--wakefield acceleration (LWFA), phase velocity is generally close to the group velocity of the laser driver and usually corresponds to a fairly small Lorentz factor in dense plasmas, so dephasing is a major limiting factor of acceleration.
For plasma--wakefield acceleration (PWFA), when a particle bunch is used as a driver, the phase velocity corresponds to the Lorentz factor of the bunch, which can be extremely large, and therefore dephasing rarely limits the acceleration length.
In this paper, we study the effects of dephasing on acceleration in radiation-dominated regime and show that the equilibrium observed for the constant acceleration rate can be achieved both in PWFA and multi-stage LWFA with dephasing.

The paper is organized as follows.
Section~\ref{sec:dynamics} presents general equations for an electron's motion in a plasma wakefield under the influence of the radiation reaction force; averaging over betatron oscillations is performed.
In Sec.~\ref{sec:radiation_dominated}, the radiation-dominated regime and the equilibrium in the uniform accelerating field is described both in the classical and the strong QED limits of radiation reaction; the applicability of the classical limit is discussed.
Section~\ref{sec:requirements} describes the parameters of the accelerated electrons required to reach the radiation-dominated regime.
In Sec.~\ref{sec:equilibrium_dephasing}, we find the conditions under which dephasing does or does not affect the equilibrium value of the radiation reaction.
In Sec.~\ref{sec:multi-stage}, we consider multi-stage acceleration with very fast dephasing during each stage and show that the same equilibrium can still be observed for values averaged over several stages.
The paper also has three appendices: Appendix~\ref{sec:quantum_corrections} further elaborates on the applicability of the classical limit, Appendix~\ref{sec:radiationless} describes the dynamics of the electron in the absence of of radiation reaction force at the initial stage of acceleration, and Appendix~\ref{sec:monte-carlo} provides the comparison to probabilistic Monte Carlo simulations of radiation reaction.

\section{Electron dynamics in a plasma wakefield with radiation reaction}
\label{sec:dynamics}

Consider a wake wave propagating along the $z$ direction with the phase velocity $v_\phase$.
Under the assumption that the wave is axisymmetric, only $E_z$, $E_r$ и $B_\phi$ components of the electromagnetic (EM) field are left.
If the evolution of the wake wave can be neglected, all these components depend on the co-propagating longitudinal coordinate $\zeta = z - v_\phase t$ and the transverse coordinate $r$.

An ultrarelativistic electron in such an EM field experiences the Lorentz force and the radiation reaction force
\begin{equation}
    \dv{\vb{p}}{t} = \vb{F}_\lorentz + \vb{F}_\rr, \quad \dv{\vb{r}}{t} = \frac{\vb{p}}{m \gamma},
\end{equation}
where
\begin{align}
    &\vb{F}_\lorentz = - e \left(\vb{E} + \frac{\vb{v}}{c} \cp \vb{B} \right), \\  
    &\vb{F}_\rr = - P(\chi) \frac{\vb{v}}{c^2}, \label{eq:LLrrforce}
\end{align}
$\gamma = (1 - \vb{v}^2/c^2)^{-1/2}$ is the Lorentz factor.
The radiation reaction force is written via a semi-classical approach \cite{Kirk_2009_PPCF_51_85008, Bulanov_2013_PRA_87_62110, Esirkepov_2015_PLA_379_2044}.

If an ultrarelativistic electron co-propagates with the wave ($p_z \gg \abs{\vb{p}_\perp}$), then $v_z \approx c$, and the forces acting on the electron can be approximated as $F_{\lorentz,z}  \approx  - e E_z$, $F_{\lorentz,r}  \approx  - e (E_r - B_\phi)$. In a plasma wakefield, the Lorentz force can be written as $F_{\lorentz, z}  =  m c \omega_\plasm f(\zeta)$, $F_{\lorentz, r}  =  - m \omega_\plasm^2 K^2 r$, where $\omega_\plasm = (4 \pi e^2 n_\plasm / m)^{1/2}$ is the electron plasma frequency, $n_\plasm$ is the unperturbed number density of electrons in the plasma, $K$ is the focusing constant.
In particular, for the strongly nonlinear (blow-out or bubble) regime of plasma wakefield, when the plasma electron density around the axis $r = 0$ is close to zero, $K^2 = 1/2$.
For quasilinear and slightly nonlinear waves, when not all electrons are expelled from the axis, the value of $K^2$ is lower.

It is more convenient to write the motion equations in dimensionless plasma units by normalizing time to $\omega_\plasm^{-1}$, spatial coordinates to $c/\omega_\plasm$, electron velocity and momentum to $c$ and $mc$.
For simplicity, we also assume that the electron has no angular motion and moves only in the $(y, z)$ plane.
In this case, the equations of motion are
\begin{align}
    &\dv{p_y}{t} = - K^2 y - \frac{p_y}{\gamma} P(\chi),  \label{eq:pyEquation} \\
    &\dv{y}{t} =  \frac{p_y}{\gamma},\\
    &\dv{p_z}{t} =  f(\zeta) - \frac{p_z}{\gamma} P(\chi), \\
    &\dv{\zeta}{t} =  \frac{p_z}{\gamma} - v_\phase, \label{eq:gammaEquation}
\end{align}
where $\gamma = \sqrt{1 + p_y^2 + p_z^2}$.
In this configuration of the EM fields, the QED parameter is calculated as
\begin{equation}
    \chi \approx \frac{\gamma \abs{F_{\lorentz,r}}}{E_\schwinger} = \frac{\gamma K^2 \abs{y}}{E_\schwinger}, \quad E_\schwinger = \frac{mc^2}{\hbar \omega_\plasm}.
\end{equation}
In the normalized units, the value of the Sauter--Schwinger field $E_\schwinger$ depends on the plasma density and is proportional to $n_\plasm^{-1/2}$.
As the dimensionless power $P(\chi)$ is also normalized to $m c^2 \omega_\plasm$, it now explicitly depends on the plasma density too.
To get the expression for this dimensionless power, we have to replace $m^2 c^4 / \hbar$ with the dimensionless $E_\schwinger$ in Eqs.~(\ref{eq:radiationReactionPower}, \ref{eq:radiationReactionPowerClassical}, \ref{eq:radiationReactionPowerQED}).
The radiation power $P(\chi)$ is the only term in system (\ref{eq:pyEquation}--\ref{eq:gammaEquation}) that contains dependence on the plasma density.

If $\gamma$ changes slowly enough and the transverse radiation reaction force is small compared to the focusing force, electrons oscillate in the transverse direction with a frequency
\begin{equation}
    \omega_\beta = \frac{K}{\sqrt{\gamma}}.
\end{equation}
We introduce new variables, the amplitude and the phase of the betatron oscillations,
\begin{align}
    &\rho^2 = y^2 + \frac{p_y^2}{K^2 \gamma}, \label{eq:rhoDefinition} \\
    &\phi = \int \omega_\beta \dd{t'},\\
    &y \approx \rho \cos \phi , \quad p_y \approx K \sqrt{\gamma} \rho \sin\phi.
\end{align}

By averaging the system of equations (\ref{eq:pyEquation}--\ref{eq:gammaEquation}) over $\phi$ and neglecting the contribution of the fast varying components, we get
\begin{align}
    &\dv{\kappa}{t} = \frac{3 \kappa}{4 \Gamma} \left[f(\xi) -  F(\kappa) \right],\label{eq:kappaEquation}\\
    &\dv{\Gamma}{t} = f(\xi) - R(\kappa),\label{gf2} \\
    &\dv{\xi}{t} = \frac{1}{2\gamma_\phase^2} - \frac{1}{2 \Gamma^2} - \frac{E_\schwinger^2 \kappa^2}{4 K^2 \Gamma^3},\label{eq:xiEquation}\\
    &F(\kappa) = \frac{1}{2\pi}\int_{0}^{2\pi}  \frac{4+ 2\sin^{2}\phi}{3} P\left(\kappa\abs{\cos\phi} \right)\dd{\phi}\label{f1}, \\
    &R(\kappa) = \frac{1}{2\pi}\int_{0}^{2\pi}P\left(\kappa \abs{\cos\phi}\right) \dd{\phi},
    \label{eq:R_definition}
\end{align}
where $\Gamma = \avg{\gamma}$ is the slowly varying component of $\gamma$, $\xi = \avg{\zeta}$ is the slowly varying component of $\zeta$, $\kappa =  \chi_\submax = \Gamma K^2 \rho / E_\schwinger$ is the amplitude of oscillations of the QED parameter $\chi$.
We also used $\avg{f(\zeta)} \approx f(\xi)$, as the difference $\sim f''(\xi) \left(\avg{\zeta^2} - \xi^2\right) / 2$ is of the second order in the small oscillation amplitude or is strictly equal to zero if the accelerating force $f(\zeta)$ is constant or linear.
As the definition of $P \propto E_\schwinger$ depends on plasma density, both $F$ and $R$ explicitly depend on plasma density as well.
The physical meaning of $R$ is that it is equal to the average rate of energy damping due to radiation reaction; the physical meaning of $F$ will be discussed later.
In the absence of the accelerating force ($f(\xi) = 0$), Eqs.~(\ref{eq:kappaEquation}--\ref{eq:R_definition}) reduce to the equations derived in Ref.~\cite{Samsonov_2021_arXiv_2107.04787}.

\begin{figure}[tb!]
    \includegraphics[width=\linewidth]{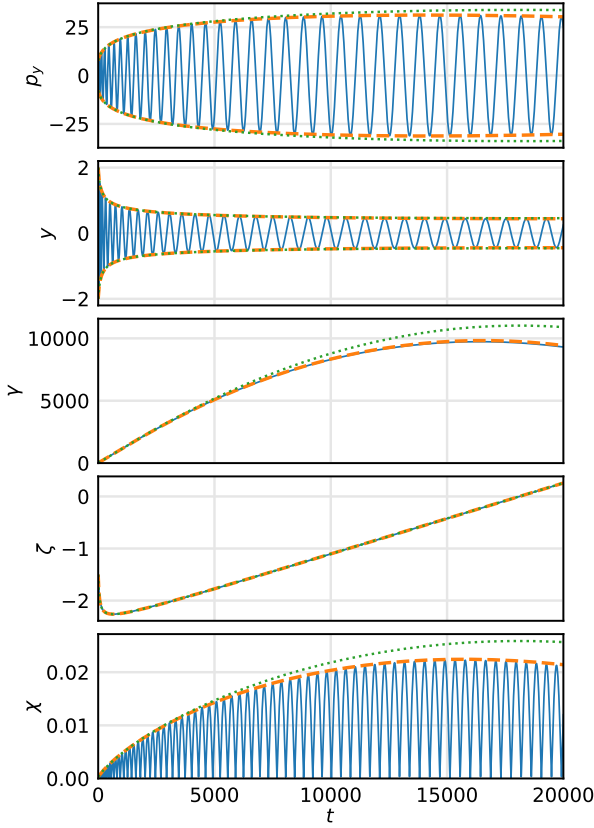}
    \caption{Dependencies $y(t)$, $p_y(t)$, $\gamma(t)$, $\zeta(t)$, and $\chi(t)$ in the full (\ref{eq:pyEquation}--\ref{eq:gammaEquation}) (solid lines) and averaged (\ref{eq:kappaEquation}--\ref{eq:xiEquation}) (dashed lines) systems.
    Initial conditions: $y_0 = 2$, $p_{y,0} = 0$, $\gamma_0 = 20$, $\zeta_0 = -1.5$.
    Wakefield parameters: $\gamma_\phase = 60$, $K^2 = 1/2$, $n_\plasm = \SI{2e22}{\cm^{-3}}$, $f(\zeta) = -\zeta / 2$.
    The dotted lines correspond to the solution of the averaged system (\ref{eq:kappaEquation}--\ref{eq:xiEquation}) without radiation reaction ($P(\chi) = 0$).
    }
    \label{fig:oscillations}
\end{figure}

Figure~\ref{fig:oscillations} shows an example of electron dynamics in a strongly nonlinear wave ($K^2 = 1/2$, $f(\zeta) = - \zeta/ 2$) according to the full (\ref{eq:pyEquation}--\ref{eq:gammaEquation}) and the averaged (\ref{eq:kappaEquation}--\ref{eq:xiEquation}) systems of equations.
As expected, the averaged equations provide the same solution, so they will be used from now on.

Now, consider the case when the slippage phase is constant ($\xi = \text{const} $) and, thus, the accelerating force is also constant ($f(\xi) = f_0 = \text{const}$). Then the system of equations has the first integral 
\begin{align}
    &\ln \Gamma  - G (\kappa )   =  \text{const} \label{integral} , \\
    &G(\kappa) =   \frac{4}{3} \int^\kappa \frac{f_0 - R(\kappa')} {f_0 - F(\kappa')} \frac{\dd{\kappa'}}{\kappa'}.
\end{align}
In the limit $f_0=0$, the first integral coincides with one for beam-beam interaction \cite{Samsonov_2021_arXiv_2107.04787}.
It should be noted that Eq.~\eqref{eq:kappaEquation} can be rewritten as follows 
\begin{equation}
   \dv{\kappa}{\tau}  = \frac{3 \kappa}{4 }\left[f_0 - F(\kappa) \right],
   \label{main}
\end{equation}
where 
\begin{equation}
    \dv{\tau}{t} = \frac{1}{ \Gamma (t)} 
\end{equation}
and $\tau$ is the proper time of the particle.
The solution can be presented in the implicit form 
\begin{equation}
    \tau = \int \frac{4 \dd{\kappa}}{ 3 \kappa \left[f_0 -  F(\kappa)  \right]}.
\end{equation}
Interestingly, in variables $\Gamma$ and $\tau$, integral~\eqref{integral} becomes the Hamilton function describing their evolution.

In the classical limit ($\chi \ll 1$) one has $F(\kappa) = (3/2) R(\kappa)= 3 P (\kappa)/4 \propto \kappa^2$
and the first integral takes a form
\begin{equation}
     \frac{\Gamma^{9/2}}{P(\kappa)^3} \left[f_0 - \frac{3}{4} P(\kappa) \right] = \text{const}.\label{integral1}
\end{equation}
In the QED regime ($\chi\gg1$) one has  
$F(\kappa)= (19/12) R (\kappa) = C_2  P(\kappa) \propto \kappa^{2/3} $, $C_2 = 19\sqrt{\pi}/[2 \GammaFunc(1/6) \GammaFunc(1/3)] \approx 1.13$
and the first integral is
\begin{equation}
   \left( \frac{\Gamma}{P^2(\kappa)} \right)^{19/14} \left[f_0 - C_2 P(\kappa) \right] = \text{const}.\label{integral2}
\end{equation}
Again, in the limit $f_0=0$, Eqs.~\eqref{integral1} and \eqref{integral2} reduce to the first integrals calculated in Ref.~\cite{Samsonov_2021_arXiv_2107.04787}.

\section{Radiation-dominated regime of acceleration}
\label{sec:radiation_dominated}

First, we analyze Eq.~\eqref{main} for the constant accelerating force. For the dynamical system governed by Eq.~\eqref{main} the only equilibrium exists when the RHS of equation is equal to $0$.
The equilibrium value of $\kappa_0$ can be found from equation
\begin{equation}
    f = F(\kappa_0), \label{eq:equilibrium_kappa_qed}
\end{equation}
So the physical meaning of function $F(\kappa)$ is that it is equal to the acceleration gradient at which the current level of the maximum QED parameter $\kappa_0 = \chi_\submax$ would remain unchanged during acceleration.

\begin{figure}[tb]
    \includegraphics[width=\linewidth]{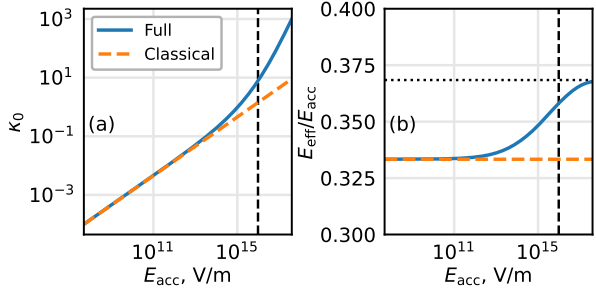}
    \caption{(a) The dependence of the equilibrium value of $\kappa_0$ and (b) the relative effective accelerated field $E_\effective/E_\accelerating$ on the accelerating field $E_\accelerating$.
    The dashed lines correspond to the solution in the classical limit.
    The vertical lines show the threshold field $\alpha E_\schwinger$, and the dotted horizontal line in (b) corresponds to the QED limit of $7/19$.
    }
    \label{fig:equilibrium_chi_acc}
\end{figure}

In order to analyze the equilibrium stability, we can expand Eq.~\eqref{main} near $\kappa_0$: 
\begin{align}
    &\kappa = \kappa_0 + \delta, \\
    &\dv{\delta}{\tau} =  - \frac{3 \kappa_0}{4} F' \left(\kappa_0 \right) \delta.
\end{align}
Since $\kappa > 0$ and $F(\kappa)$ is a positive monotonically increasing function (by definition, because $P(\chi)$ is also a monotonically increasing function), the equilibrium point $\kappa = \kappa_0$ is stable.
Regardless of the initial conditions, it is approached exponentially but never reached.
Also, as $\Gamma$ cannot grow faster than linearly with $t$, $\tau \to +\infty$ corresponds to $t \to + \infty$, so this equilibrium will always be approached in the true time $t$ as well.
Therefore, in the limit $t \to + \infty$, the averaged electron trajectory is attracted to the trajectory with constant $\kappa = \kappa_0$ corresponding to the radiation-dominated regime of acceleration.
If we return to physical units, we can show that the equilibrium condition does not depend on plasma density,
\begin{equation}
    \frac{E_\accelerating}{\alpha E_\schwinger} =  \frac{1}{3\pi} \int_0^{2\pi} (2 + \sin^2\phi) \frac{\hbar P(\kappa_0 \abs{\cos\phi})}{\alpha m^2 c^4} \dd{\phi}.
\end{equation}
Therefore, the equilibrium value $\kappa_0$ corresponding to the maximum value of QED parameter $\chi_\submax$ is determined only by the accelerating gradient of the accelerator (see Fig.~\ref{fig:equilibrium_chi_acc}a).
This applies to any kind of accelerator with a linear focusing force.

The rate of acceleration is described by Eq.~\eqref{gf2}.
At the equilibrium point $\kappa = \kappa_0$,
\begin{equation}
    \dv{\Gamma}{t} = f - R\left(\kappa_0 \right) = f\left[1 - \frac{R(\kappa_0)}{F(\kappa_0)} \right].
\end{equation}
Due to the definition of functions $R$ and $F$ and the fact that $P(\chi)$ is monotonously increasing, the effective accelerating force lies between $f/4$ and $f/2$, so radiation reaction cannot prevent infinite acceleration.
As the effective force is constant, the particle energy increases linearly in time in the radiation-dominated regime.
In the classical limit ($\chi \ll  1$), $\Gamma = (1/3) f t$, so the acceleration rate in radiation-dominated regime is reduced by a factor of $3$ compared to the regime without radiative losses (see Ref.~\cite{Kostyukov_2012_PRSTAB_15_111001}).
The equilibrium value of $\kappa_0$ is equal to 
\begin{equation}
   \kappa_0 = \chi_\submax = \sqrt{2 \frac{E_{acc}}{\alpha E_\schwinger}}.
   \label{eq:chi_max}
\end{equation}
The transition to the QED regime occurs at very high value of the accelerated field $E_\accelerating \sim \alpha E_\schwinger$, or approximately \SI{e16}{V/m}, which is far beyond the capabilities of any plasma accelerators.
Plasma accelerators can provide accelerating gradients comparable to the plasma wavebreaking limit $E_\plasm[\si{V/m}] \approx 96 \sqrt{n_\plasm[\si{\cm^{-3}}]}$, which can only reach \si{TV/m} levels of acceleration even for solid-state density plasmas.
If the QED limit ($\chi \gg  1$) is reached in some other type of accelerator with a linear focusing force, then $\Gamma = (7/19) F t$, and the acceleration rate is slightly higher than in the classical limit (see Fig.~\ref{fig:equilibrium_chi_acc}b).

So, the classical limit of radiation reaction can always be used for plasma accelerators with reasonable parameters (see Appendix~\ref{sec:quantum_corrections} for more details on the limiting cases).
In this case, the QED parameter $\chi \ll 1$ and the corresponding variable $\kappa$ have low importance, and it is more convenient to rewrite the radiation reaction power through the oscillation amplitude,
\begin{equation}
    P = \epsilon K^4 \rho^2 \Gamma^2, \quad \epsilon = \frac{2}{3} \frac{\omega_\plasm}{c} \frac{e^2}{mc^2} \propto \sqrt{n_\plasm}.
\end{equation}
The value of $\epsilon = 1$ is reached for $n_\plasm \approx \SI{8e36}{\cm^{-3}}$, way above even solid-state densities, so in plasma accelerators $\epsilon \ll 1$.
The functions $R$ and $F$ in this case are
\begin{equation}
    R = \frac{1}{2} P, \quad F = \frac{3}{4} P = \frac{3}{2} R.
\end{equation}
Correspondingly, as $\kappa$ is no longer a parameter in $P$, it is convenient to use the average radiation damping rate $R$ as the new variable and rewrite the system of equations as
\begin{align}
    &\dv{R}{t} = \frac{9}{4} \frac{R}{\Gamma} \left(\frac{2 f(\xi)}{3} - R \right), \label{eq:fRREquation}\\
    &\dv{\Gamma}{t} = f(\xi) - R, \label{eq:gammaEquationFRR}\\
    &\dv{\xi}{t} = \frac{1}{2\gamma_\phase^2} - \frac{1}{2\Gamma^2} - \frac{R}{2 \epsilon K^2 \Gamma^3}. \label{eq:zetaEquationFRR}
\end{align}

As already stated, when $f(\xi) = f_0 = \text{const}$, then $R$ eventually reaches an equilibrium value of $2f_0/3$, and acceleration is slowed by a factor of $3$ but never stops.
However, when the accelerating gradient $f(\xi)$ is not constant, $R = 2f (\xi) / 3$ may no longer be an equilibrium if $f(\xi)$ changes fast enough due to the electron slipping along the $\xi$ coordinate.
In this paper, we analyze the above system and find the conditions under which acceleration can reach the equilibrium $R = 2 f /3$ and whether it is possible to reach it in realistic laser-- or plasma--wakefield accelerators.

\section{Requirements for radiation-dominated regime}
\label{sec:requirements}

\begin{figure}[tb]
    \includegraphics[width=\linewidth]{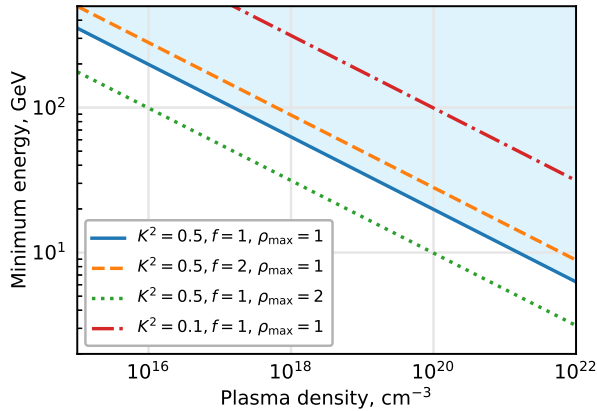}
    \caption{The dependence of the minimum energy above which the oscillation amplitude $\rho$ required to reach the radiation-dominated regime is lower than $\rho_\submax$ for different parameters of $K^2$, $f$, and $\rho_\submax$. 
    The shaded area shows the region where the condition $\rho < \rho_\submax$ is satisfied for the first set of parameters.}
    \label{fig:rr_energy_vs_density}
\end{figure}

In order to be in the radiation-dominated regime, we must satisfy the condition
\begin{equation}
    R = \frac{1}{2} \epsilon K^4 \rho^2 \Gamma^2 = \frac{2f}{3},
\end{equation}
so the amplitude of the oscillations should be equal to
\begin{equation}
    \rho = \sqrt{\frac{4 f}{3 \epsilon K^4 \Gamma^2}}.
\end{equation}
The amplitude $\rho$ (which is equal to the maximum deviation from the axis $y_\submax$) cannot be very large, as it is limited by the transverse size of the wake (typically of the order of unity).
So, we must fulfill the condition $\rho < \rho_\submax$, or
\begin{equation}
    \Gamma > \Gamma_\submin = \sqrt{\frac{4 f}{3\epsilon K^4 \rho_\submax^2}} \propto n_\plasm^{-1/4}.
    \label{eq:rr_regime_condition}
\end{equation}
Both $f$ and $\rho_\submax$ are typically of the order of unity, and $K^2$ is of the order of $1/2$, so the right-hand side is determined mostly by $\epsilon$ which depends on the plasma density.
Figure~\ref{fig:rr_energy_vs_density} shows the dependence of the minimum energy of electrons required to reach the radiation-dominated regime.
For typical plasmas used in plasma accelerators, this energy is of the order of 10 to \SI{100}{\GeV}.
However, the use of weaker focusing gradients $K^2$ corresponding to quasilinear waves significantly increases the required energy.

Of course, as the electron overtakes the wave, it may reach the area where the accelerating gradient $f$ is close to $0$, and thus satisfying condition \eqref{eq:rr_regime_condition} becomes easy.
However, if the contribution of this region to the overall energy gain is small, we cannot consider this regime radiation-dominated despite formally satisfying the condition.

Another consideration is that not only the final, but the initial amplitude of betatron oscillations $\rho_0$ should not be larger than $\rho_\submax$ either.
If we start from low energies in the radiationless regime $R \ll f$, then the oscillation amplitude $\rho$ quickly decreases with acceleration and, by the time we reach the radiation-dominated regime, $\rho \ll \rho_0$.
This makes the condition for reaching the radiation-dominated regime stricter.
The full description of the electron dynamics in the radiationless regime can be found in Appendix~\ref{sec:radiationless}.
Here, we only use that in the radiationless limit $R$ scales as
\begin{equation}
    R = R_0 \qty(\frac{\Gamma}{\Gamma_0})^{3/2}
\end{equation}
regardless of the longitudinal motion of the particle.
Therefore, the amplitude $\rho$ scales as 
\begin{equation}
    \rho = \rho_0 \qty(\frac{\Gamma_0}{\Gamma})^{1/4},
\end{equation}
which is a well-known result for betatron oscillations in the absence of radiation reaction \cite{Kostyukov_2004_PoP_11_5256}.
Assuming that this scaling approximately holds by the time the radiation dominated regime $R = 2f/3$ is reached, the demand that $\rho_0 < \rho_\submax$ yields a different condition for $\Gamma$,
\begin{equation}
    \Gamma > \Gamma_\submin \qty(\frac{\Gamma_\submin}{\Gamma_0})^{1/3},
\end{equation}
where $\Gamma_\submin$ is the same as in Eq.~\eqref{eq:rr_regime_condition}.
For example, if we use a $\SI{e17}{\cm^{-3}}$ plasma with $\Gamma_\submin \sim \SI{100}{\GeV}$, and the initial energy of the bunch is $\Gamma_0 = \SI{100}{\MeV}$, then we need to accelerate the bunch to the energy $\Gamma > \SI{1000}{\GeV}$ to possibly reach the radiation-dominated regime.

The required energies can be reached either in a very long single-stage accelerator driven by a particle bunch, when dephasing happens very slowly, or after many shorter acceleration stages.
Of course, if the transverse quality of the bunch degrades between the stages (i.e., the oscillation amplitude increases after the transition from one stage to another), this regime can be reached sooner, but this is usually unwanted.

\section{Radiation reaction force equilibrium with dephasing}
\label{sec:equilibrium_dephasing}

If we reach the radiation dominated regime, then the growth of $R$ saturates at the value $R = 2 f/3$ if the value of $f(\xi)$ is constant or changes slowly enough.
Let us find the condition when this is true.
Assuming that the electron's own contribution to phase slippage can be neglected (i.e., $1 - v_z \ll 1 - v_\phase$), the time evolution of $f$ can be found as
\begin{equation}
    \dv{f}{t} = \dv{f}{\xi} \dv{\xi}{t} = \frac{f'}{2 \gamma_\phase^2},
\end{equation}
where $f' = -\dv*{f}{\xi}$.
For a strongly nonlinear wake wave, $f' = -1/2$ in most parts of the wake.

Let us find the condition when this change is too fast for the equilibrium to be set.
To do so, we introduce $R_\relative = R / f$, and write the equation for $R_\relative$,
\begin{equation}
    \dv{R_\relative}{t} = \frac{9}{4} \frac{R_\relative}{\Gamma} \left(\frac{2}{3} - R_\relative  \right) - \frac{R_\relative f'}{2 f^2 \gamma_\phase^2}.
\end{equation}
The ``equilibrium'' for $R_\relative$ is 
\begin{equation}
    R_\relative = \frac{2}{3} - \frac{2\Gamma f'}{9 f^2 \gamma_\phase^2}.
    \label{eq:Requilibrium}
\end{equation}
Usually, $f'<0$, and thus the second term is a positive correction.
Much like the value $2f/3$ for $R$, it is not a true equilibrium, as its value changes with time.
However, the condition when the value of $R_\relative$ remains close to $2/3$ is 
\begin{equation}
    \frac{\Gamma \abs{f'}}{f^2 \gamma_\phase^2} \ll 1.
    \label{eq:equilibriumCondition}
\end{equation}
The physical meaning of this condition is pretty simple: the typical time of change in the value of $f$ (equal to $2\gamma_\phase^2 f/ \abs{f'}$) should be much larger then the typical time of approaching the equilibrium in Eq.~\eqref{eq:fRREquation} (equal to $R/\gamma \sim f/\gamma$).
As $f$ decreases to $0$ and $\Gamma$ increases when the electron is accelerated, this condition will often eventually break.
However, it might be fulfilled for a large part of acceleration.
Assuming the accelerating force $f \sim 1$, and $\abs{f'} \sim 1$, the necessary condition is
\begin{equation}
    \Gamma \ll \gamma_\phase^2.
\end{equation}
As the energy gained over the entire accelerating stage is proportional to $\gamma_\phase^2$ (see Appendix~\ref{sec:radiationless}), such a condition can only be fulfilled for a single-stage case.
So, for LWFA, getting $R \approx 2/3$ is basically not possible at all, either due to not being able to reach the radiation-dominated regime in the case of single-stage acceleration, or due to the rate of dephasing being too high to satisfy condition~\eqref{eq:equilibriumCondition}.
However, in a single-stage PWFA, this condition can be fulfilled.
In this case, even though Eq.~\eqref{eq:Requilibrium} is not a true solution, it can serve as an estimate for how much $R_\relative$ deviates from $2/3$ when this deviation is small enough.

\begin{figure}[tb]
    \includegraphics[width=\linewidth]{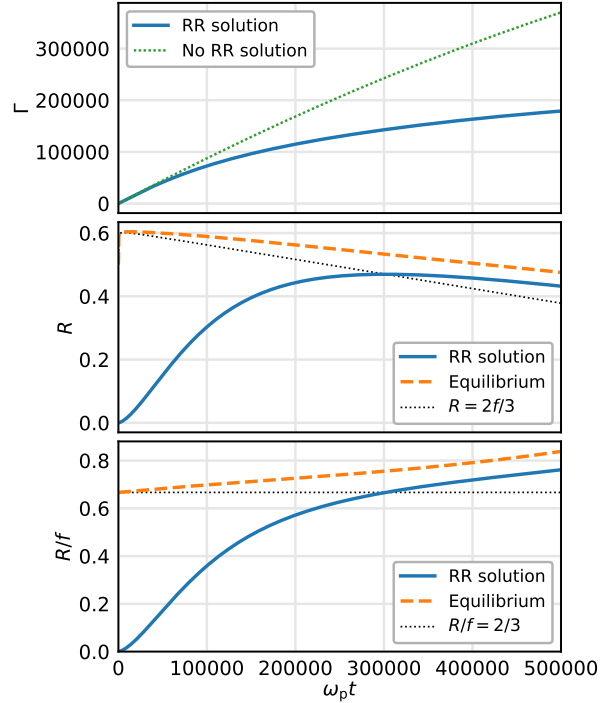}
    \caption{The numerical dependencies of $\Gamma$, $R$, and $R_\relative=R / f$ on time $t$.
    The dotted line in the top picture corresponds to the solution without radiation reaction.
    The dashed line in the other two pictures shows the instantaneous equilibrium \eqref{eq:Requilibrium}.
    The dotted line shows the level $R = 2f/3$ expected for $f = \text{const}$.
    Properties of the wake: $n_\plasm = \SI{e22}{cm^{-3}}$, $\gamma_\phase =600$, $K^2 = 1/2$, $f(\xi) = - \xi /2$.
    Initial parameters of the electron: $\Gamma_0 = 20$, $\xi_0 = -1.5$, $\rho_0 = 1$.
    }
    \label{fig:equilibrium}
\end{figure}

Figure~\ref{fig:equilibrium} shows an example for single-stage PWFA when the saturation of the radiation reaction force at the level of almost $2/3$ is achieved.
We see that $R$ reaches values comparable to the accelerating force $f$, but stops growing with $\Gamma$ and saturates, thus not preventing the electron from further acceleration.
Then it even begins to decline with the decrease in the accelerating force $f$ due to dephasing.
At the same time, the ratio $R_\effective$ stays between the value of $2/3$ and the equilibrium value predicted by Eq.~\eqref{eq:Requilibrium}.

\section{Multi-stage acceleration}
\label{sec:multi-stage}

Now consider a situation when $\gamma_\phase$ is pretty low, so the electron rapidly dephases, and acceleration consists of many stages.
In this case, condition \eqref{eq:equilibriumCondition} is unlikely to be satisfied, which is typical for the LWFA case.
However, if every acceleration stage is short, we might introduce values averaged over stages,
\begin{equation}
    R = \avg{R} + \delta R, \quad \Gamma = \avg{\Gamma} + \delta \Gamma,
\end{equation}
where $\delta R$ and $\delta \Gamma$ change fast within one stage, and $\avg{R}$ and $\avg{\Gamma}$ change slowly compared to the duration of one stage.
This is a secondary averaging, as $R$ and $\Gamma$ are already values averaged over betatron oscillations. Therefore, each stage has to be much longer than the betatron oscillation period to make such separation possible.
The condition of applicability of this consecutive averaging is thus
\begin{equation}
    \frac{2\pi}{\omega_\beta} \ll 2 \gamma_\phase^2 L, \quad \Gamma \ll \frac{\gamma_\phase^4 L^2 K^2}{\pi^2}.
\end{equation}
where $L$ is the length of the accelerating phase (i.e., the total slippage of the electron in the co-moving coordinate $\xi$).
As the maximum energy scales as $\gamma_\phase^4$, it can usually be satisfied.
If not, full equations of motion should be used instead of the averaged equations.

We consider only identical acceleration stages in which the plasma density, the phase velocity, and the parameter $K$ are exactly the same.
We also assume that the electrons can be transported between the stages without the change in their energy and oscillation amplitude, so $\Gamma$ and $R$ remain continuous between the stages.
We also only count the time spent inside the stages, assuming that the transition between the stages happens instantaneously.
From the point of view of Eqs.~(\ref{eq:fRREquation}--\ref{eq:gammaEquationFRR}), this corresponds to replacing $f(\xi)$ with $f(t, \xi)$, where $f(t, \xi)$ is a function whose time dependence correspond to abrupt changes in profile between each stages.

If every accelerating stage is sufficiently short, we can assume that $\abs{\delta R} \ll \avg{R}$, $\abs{\delta \Gamma} \ll \avg{\Gamma}$.
At the same time, $\delta f = f - \avg{f}$ cannot be considered small, as the variation of the accelerating force over one section can be of the same amplitude as the force, e.g. when we end each stage when the electron reaches the point of no acceleration ($f = 0$).
This allows us to linearize Eqs.~(\ref{eq:fRREquation}--\ref{eq:gammaEquationFRR}) using the smallness of $\delta R$ and write separate equations for the averaged and rapidly oscillating values,
\begin{align}
    &\dv{\avg{R}}{t} = \frac{9}{4} \frac{\avg{R}}{\avg{\Gamma}} \left(\frac{2 \avg{f}}{3} - \avg{R} \right),\\
    &\dv{\avg{\Gamma}}{t} = \avg{f} - \avg{R},\\
    &\dv{\delta R}{t} = \frac{3}{2} \frac{\avg{R}}{\avg{\Gamma}} \delta f, \quad \dv{\delta \Gamma}{t} = \delta f. \label{eq:deltaRdeltaGamma}
\end{align}
As the integral of $\delta f$ over one stage is equal to $0$, $\delta R$ and $\delta \gamma$ only oscillate around $0$ from stage to stage, which retroactively confirms our assumptions.
The equations for averaged $\avg{R}$ and $\avg{\Gamma}$ are exactly the same as initial equations~(\ref{eq:fRREquation}--\ref{eq:gammaEquationFRR}), but with the averaged accelerating force $\avg{f}$ instead.

The equations above did not include the motion along the longitudinal coordinate $\xi$.
If we now have a fast enough electron that it moves with the velocity $\dv*{\xi}{t} = 1/(2\gamma_\phase^2)$ and all stages are identical, then the average force can be calculated as
\begin{equation}
    \avg{f} = \frac{1}{\xi_1 - \xi_0} \int_{\xi_0}^{\xi_1} f(\xi) \dd{\xi},
\end{equation}
where 
\begin{equation}
    \xi_1 = \xi_0 + L = \xi_0 + \frac{T}{2\gamma_\phase^2},
\end{equation}
where $T$ is the duration of the stage.
Assuming that we reset electrons from $\xi_1$ exactly to their initial coordinate $\xi_0$ after every stage, $\avg{f} = \textup{const}$.
This means that the evolution of the average energy $\avg{\Gamma}$ and the average radiation reaction force $\avg{R}$ follows the predictions of Sections~\ref{sec:dynamics} and \ref{sec:radiation_dominated} as well as Ref.~\cite{Kostyukov_2012_PRSTAB_15_111001} where acceleration in constant force was considered.
This includes the consequence that $\avg{R}$ will always tend to $2\avg{f}/3$ given enough time, and acceleration will never be stopped by radiation reaction.

\begin{figure}[tb]
    \includegraphics[width=\linewidth]{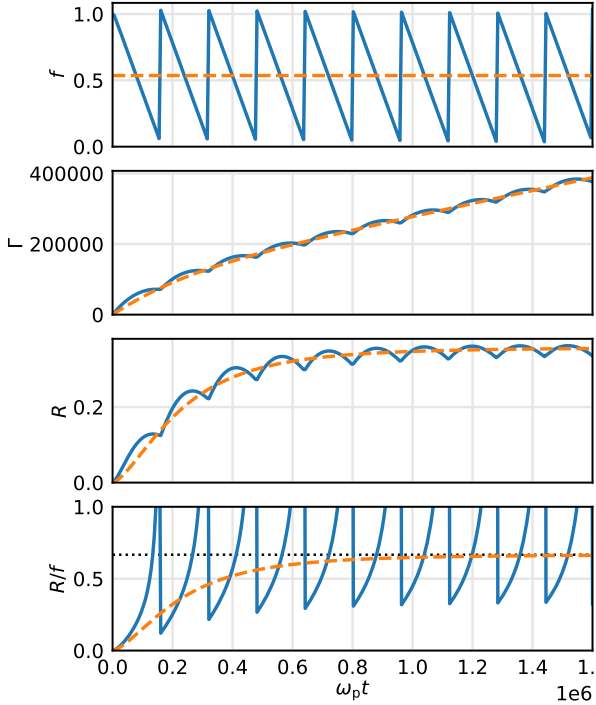}
    \caption{Time dependencies of $f$, $\Gamma$, $R$ and $R/f$ in a multistage accelerator with $n_\plasm = \SI{e21}{cm^{-3}}$, $\gamma_\phase = 200$, $K^2 = 1/2$ calculated numerically by solving Eqs.~(\ref{eq:fRREquation}--\ref{eq:zetaEquationFRR}).
    The phase of the wake is shifted back by $1$ every $\omega_\plasm T = 4 \gamma_\phase^2$, corresponding to a complete restoration of phase of electrons that are much faster than the wave.
    Initial parameters of the electron are $\Gamma_0 = 500$, $\xi_0 = -2$, $\rho_0 = 0.5$.
    The average value of the accelerating force was equal to $\avg{f} \approx 0.54$.
    The dashed lines correspond to constant $f(\xi) = \avg{f} = 0.54$ with the same initial parameters of electron.}
    \label{fig:multi_stage}
\end{figure}

Figure~\ref{fig:multi_stage} shows an example of multi-stage acceleration dynamics compared to acceleration under the influence of a constant force.
It demonstrates that, on average, multistage acceleration can reduced to acceleration in a constant average field despite very fast dephasing.
And even though condition~\eqref{eq:equilibriumCondition} is not satisfied, and the instantaneous value of $R/f$ is rapidly oscillating, the averaged over stages radiation reaction force $\avg{R}$ reaches the same equilibrium value of $2/3$ of the average accelerating force.

We can also estimate the amplitude of oscillations $\delta \gamma$ and $\delta R$.
Do to so, we replace $\dv*{t} = 1 / (2\gamma_\phase^2) \dv*{\xi}$ in Eq.~\eqref{eq:deltaRdeltaGamma}, and assume that $\delta f$ is linear,
\begin{equation}
    \delta f = - \abs{f'} \left(\xi - \frac{\xi_0 + \xi_1}{2}\right).
\end{equation}
In this case, the amplitudes of oscillations are
\begin{align}
    &\delta \gamma_\submax - \delta\gamma_\submin = \frac{1}{4} \gamma_\phase^2 L^2 \abs{f'},\\
    &\delta R_\submax - \delta R_\submin = \frac{3}{8} \frac{\gamma_\phase^2 L^2 \abs{f'}}{\avg{\Gamma}} \avg{R}.
\end{align}
To satisfy the conditions of applicability of this averaged approach, we need $\delta R \ll \avg{R}$.
The highest amplitude of oscillations is achieved when the slope of accelerating force $\abs{f'}$ is the largest.
Assuming the accelerating force is always positive, the largest value of $\abs{f'}$ is equal to $2\avg{f}/L$, which correspond to $f$ going from $2\avg{f}$ to $0$ over the accelerating stage.
Therefore, to satisfy the condition of applicability, we need
\begin{equation}
    \gamma_\phase^2 L \avg{f} \ll \avg{\Gamma}.
\end{equation}
This condition is basically the opposite to condition \eqref{eq:equilibriumCondition} for when the equilibrium $R = 2f/3$ is achieved.
So, in this regime, even though the averaged values satisfy the equilibrium condition $\avg{R} = 2\avg{f}/3$, equilibrium is not observed for non-averaged values, as evidenced by Fig.~\ref{fig:multi_stage}.

If we consider an actual bunch accelerated in a multi-stage scheme, electrons starting from different initial coordinates will have different average accelerating forces acting on them.
However, radiation damping for every electron will still be limited by condition $\avg{R} \leq 2\avg{f}/3$.
Therefore, even in multi-stage acceleration with very fast dephasing, radiation reaction still cannot prevent infinite acceleration of electrons.

\section*{Discussion and conclusions}

In this paper, we have shown that the radiation reaction generally cannot prevent infinite acceleration of electrons in plasma accelerators.
Both for the classical and strong QED radiation reaction forces, the growth of the QED parameter $\chi$ during acceleration is not infinite and reaches the equilibrium value determined solely by the amplitude of the accelerating force.
At the same time, the averaged over betatron oscillations radiation reaction force saturates at the level smaller than the accelerating force.
In the classical limit, this level is $2/3$ of the accelerating gradient, and remains close to this value even when the classical limit breaks.
Dephasing can prevent the electron from reaching the equilibrium, as the time required to reach it can become higher than the characteristic time of dephasing and the corresponding change in the accelerating gradient.
However, there are two important limiting cases when equilibrium is still observed.
For extremely slow dephasing typical for PWFA with extremely high $\gamma_\phase$ corresponding to the phase velocity very close to the speed of light, the equilibrium is almost reached for the instantaneous value of the accelerating field acting on the particle.
However, even in the opposite case of multi-stage LWFA with low phase velocity and small $\gamma_\phase$, when dephasing happens extremely fast and the instantaneous radiation reaction force cannot reach the equilibrium within any single stage, the values averaged over multiple acceleration stages still satisfy the equilibrium condition.

One limitation we used in the paper is that the focussing constant $K^2$ (unlike the accelerating force $f(\zeta)$) is assumed to be constant along the wake, so its possible change is not contributing to the effects of the electron dephasing.
This assumption is strictly true for a strongly nonlinear wake with full evacuation of electrons from the axis, when $K^2 = 1/2$ in the entire accelerating cavity.
For a slightly nonlinear wake, there is almost full evacuation of electrons, and the change of $K^2$ is also small.
But in a quasilinear wake, the focussing constant $K^2 \propto \sin(\omega_\plasm \zeta / c)$, and if we use a significant portion of the accelerating phase (with the length equal to $\lambda_\plasm / 4$), the change in $K^2$ can become significant.
Typically, a constant corresponding to value of $K^2$ averaged over the entire accelerating phase is used in theoretical studies in this case \cite{Michel_2006_PRE_74_26501, Deng_2012_PRSTAB_15_81303}.
The effect of dephasing when $K$ is non-constant can be studied by deriving the averaged system of equations in this case as well (derivatives $\dv*{K}{\zeta}$ which appear from the updated averaging procedure can be elimitated by introducing a new variable $\tilde\kappa = \kappa / K^{3/2}$), but measuring its impact on acceleration is outside the scope of this work.

Another limitation is our description of radiation reaction.
The classical radiation reaction force \eqref{eq:LLrrforce} assumes that radiation is a continuous process.
From the quantum point of view, this corresponds to the case when the electron radiates many low-energy photons, and the solution remains very close to the case described by the continuously acting force.
In the quantum limit, the energy of each radiated photon might become comparable to the energy of the electron, and the radiation process becomes inherently stochastic.
In this case, the classical description can sometimes describe only the average behavior.
Numerical methods better suited for describing such processes are Fokker--Plank stochastic simulations \cite{Niel_2018_PRE_97_43209} which describes both the average force and diffusion effects caused by its stochasticity and Monte Carlo simulations which treat the radiation of individual photons probabilistically \cite{Duclous_2010_PPCF_53_15009}.
However, as our estimates show (see Sec.~\ref{sec:radiation_dominated} and Appendix~\ref{sec:quantum_corrections}), the value of $\chi$ is of the order of $0.1$ even for the solid-state density plasmas, and therefore we can rely on the classical description in all realistic cases.
The comparison between the classical radiation reaction approach and Monte Carlo simulations for the extreme case is made in Appendix~\ref{sec:monte-carlo}.
Even then it shows fairly good agreement for the averaged values of the acceleration.
And for less dense plasmas, the effects of stochasticity will play a much less significant role.

In our work, we also focused on the single particle approach.
As the Coulomb self-force of the ultrarelativistic bunch is small and scales as $\gamma^{-2}$, these calculations remain valid for a bunch with the electron density $n_\bunch \ll \gamma^2 n_\plasm$.
However, the study of the parameters of the bunch, such as transverse emittance or energy spread, was outside the scope of this paper.
Some consequences of the radiation reaction on bunch parameters have been studied in previous works \cite{Michel_2006_PRE_74_26501, Deng_2012_PRSTAB_15_81303}.
For instance, it was shown that the radiation reaction can cause the decrease in the transverse emittance of the bunch.
In the saturated regime of acceleration, when the effective accelerating force is three times lower than the accelerating gradient, the radiation reaction force does not cause additional growth of of the energy spread compared to the radiationless case.
However, the transition to this	regime happens sooner for electrons with the largest amplitude of betatron oscillations, while electrons with smaller oscilations will continue to accelerate without being affected by radiation reaction as much, which could cause the increase in the energy spread of the bunch.
Still, this effect can be offset by the fact that electrons with large oscillations have a smaller longitudinal velocity and thus stay in larger accelerating field longer, so further research is required.

\begin{acknowledgements}

The work was supported by the Russian Science Foundation (Grant No.~20-12-00077, strong QED limit of acceleration) and the Russian Foundation for Basic Research (project No.~20-21-00150, analysis of the system in the classical limit)

We are grateful to Alexander Samsonov for valuable discussions.

\end{acknowledgements}

\appendix

\section{Classical limit applicability}
\label{sec:quantum_corrections}

\begin{figure}[tb]
    \includegraphics[width=\linewidth]{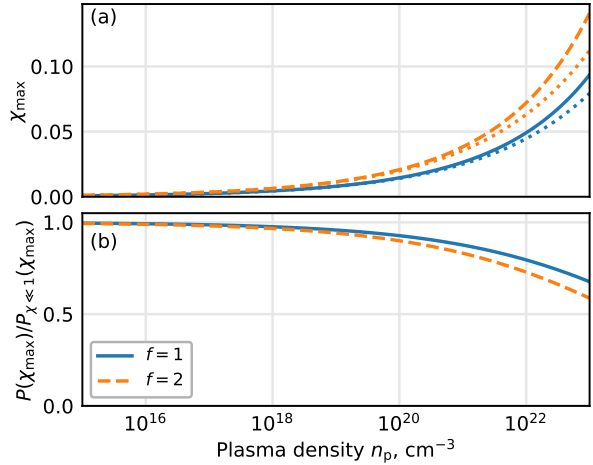}
    \caption{(a) Dependencies of $\chi_\submax$ on plasma density and (b) the corresponding values of $P(\chi_\submax) / P_{\chi \ll 1}(\chi_\submax)$ for the value of $\chi_\submax$ calculated according to the quantum formula \eqref{eq:equilibrium_kappa_qed}.
    The dotted lines in (a) show the solution \eqref{eq:chi_max_plasma} in the classical limit.
    }
    \label{fig:quantum_corr}
\end{figure}

As shown in Sec.~\ref{sec:radiation_dominated}, the maximum value of $\chi_\submax$ in the radiation-dominated regime for infinitely long acceleration will tend to the equilibrium value given by Eq.~\eqref{eq:chi_max} in the classical limit.
In plasma accelerators, where the accelerating force is $E_\accelerating = f_0 E_\plasm = f m c \omega_\plasm/e$, this formula can be rewritten as
\begin{equation}
    \chi_\submax = \sqrt{2 f_0 \frac{m c \omega_\plasm}{\alpha E_\schwinger e}}.
    \label{eq:chi_max_plasma}
\end{equation}
This value increases with the plasma density as $n_\plasm^{1/4}$, and the factor $f_0$ is typically several times larger than $1$ for strongly nonlinear wakefields and smaller than $1$ for quasilinear plasma waves.
Plasmas are generally limited by densities $\sim \SI{e23}{\cm^{-3}}$.
The dependence of $\chi_\submax$ on plasma density both according to the classical formula and the more accurate solution \eqref{eq:equilibrium_kappa_qed} is shown in Fig.~\ref{fig:quantum_corr}a.
For very dense plasmas, the value of $\chi_\submax$ can reach values $\sim 0.1$, and the classical formula and the more accurate full formula give slightly different estimates for the equilibrium value of $\chi_\submax$.
For such densities, the difference between radiation losses $P$ given by Eq.~\eqref{eq:radiationReactionPower} and the losses in the classical limit $P_{\chi \ll 1}$ given by Eq.~\eqref{eq:radiationReactionPowerClassical} can also become noticeable (Fig.~\ref{fig:quantum_corr}b).

\begin{figure}[tb]
    \includegraphics[width=\linewidth]{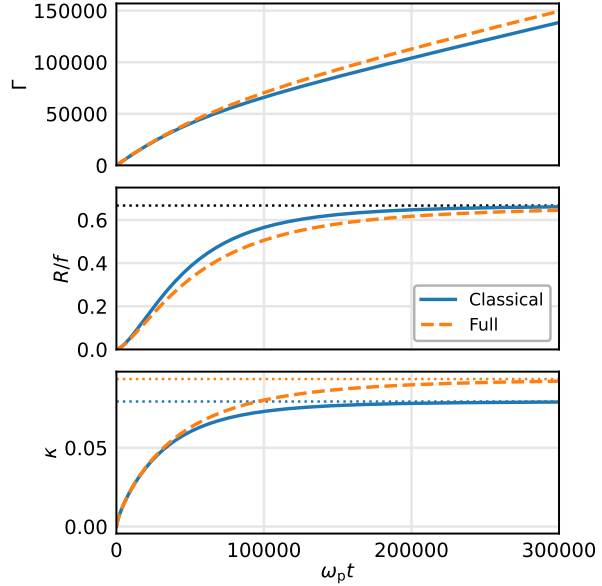}
    \caption{The dependencies of $\Gamma$, the ratio between the averaged radiation reaction force $R$ and acceleration force $f$ as well as the value of $\kappa = \chi_\submax$ on time for the classical limit of radiation reaction \eqref{eq:radiationReactionPowerClassical} (solid lines) and the full formula of the radiation reaction \eqref{eq:radiationReactionPower}.
    Thin dotted lines show the level $2/3$ for the $R/f$ ratio as well as the equilibrium values $\kappa_0$, calculated according to \eqref{eq:chi_max_plasma} and \eqref{eq:equilibrium_kappa_qed} for classical limit and the full formula, respectively.
    Wakefield parameters: $n_\plasm = \SI{e23}{\cm^{-3}}$, $\gamma_\phase = 200$, $K^2 = 1/2$, $f(\xi) = -\xi/2$.
    Initial conditions: $\Gamma_0 = 20$, $\rho_0 = 1$, $\xi_0 = -2$.
    }
    \label{fig:quantum_correction_dynamics}
\end{figure}

To see whether this difference affects the acceleration process in any significant way, we solve the system of averaged equations (\ref{eq:kappaEquation}--\ref{eq:xiEquation}) using either the classical limit of the radiation reaction power \eqref{eq:radiationReactionPowerClassical} or the full (non-classical) expression \eqref{eq:radiationReactionPower} for a plasma with the density \SI{e23}{\cm^{-3}}, for which this effect is more pronounced (see Fig.~\ref{fig:quantum_correction_dynamics}).
As already predicted in Sec.~\ref{sec:radiation_dominated} (see Fig.~\ref{fig:rr_energy_vs_density}b), the equilibrium value of the radiation reaction force remains close to $2/3$ up to extremely high acceleration gradients, although saturation happens slightly slower for the non-classical formula.
Thus, the time dependence $\Gamma(t)$ remains virtually the same.
In both cases, the value of $\kappa = \chi_\submax$ also reaches the equilibrium value predicted in Sec.~\ref{sec:radiation_dominated} (Fig.~\ref{fig:rr_energy_vs_density}a).
The difference between the classical limit and the non-classical formula for $\kappa$ are more noticeable, but they do not affect the acceleration much.
Therefore, in all realistic scenarios of plasma accelerators, even for very dense plasmas, the classical limit of radiation reaction can safely be used.

Of course, higher values $\chi > \chi_\submax$ can be possibly achieved by injecting an ultrarelativistic bunch with a large transverse size into a dense plasma.
In this case, using the non-classical expression for the radiation power might become important if the initial radiation reaction force is far above the equilibrium value of $2f/3$.
However, the initial high value of $\chi$ will rapidly decrease to $\chi_\submax$, and $R$ will become equal to $2f/3$, after which the classical limit will again become applicable.
Using the classical formula from the beginning will only change the characteristic time of damping to the equilibrium.

\section{Radiationless acceleration}
\label{sec:radiationless}

At the initial stage of acceleration, the radiation reaction force is often very small, and we can use $R \ll f$ and write equations (\ref{eq:fRREquation}--\ref{eq:zetaEquationFRR}) without the influence of radiation reaction (although it still remains one of the variables),
\begin{align}
    &\dv{R}{t} = \frac{3}{2} \frac{R f(\xi)}{\Gamma},\\
    &\dv{\Gamma}{t} = f(\xi),\\
    &\dv{\xi}{t} = \frac{1}{2\gamma_\phase^2} - \frac{1}{2\Gamma^2} - \frac{R}{2 \epsilon K^2 \Gamma^3}.
\end{align}
This system of equations corresponds to the regular betatron oscillations in the absence of radiation reaction which have been studied in previous works \cite{Kostyukov_2003_PoP_10_4818, Kostyukov_2004_PoP_11_5256}, but written in an unorthodox way, so we find it convenient to describe the solution in the new variables.
From the first two equations we can get that
\begin{equation}
    R(t) = R_0 \qty(\frac{\Gamma(t)}{\Gamma_0})^{3/2},
\end{equation}
so the radiation reaction force scales as $\Gamma^{3/2}$ regardless of the dependence of the accelerating force on time or the longitudinal coordinate.
This allows us to get rid of the first equation,
\begin{align}
    &\dv{\Gamma}{t} = f(\xi), \label{eq:gammaEqNoRR}\\
    &\dv{\xi}{t} = \frac{1}{2\gamma_\phase^2} - \frac{1}{2\Gamma^2} - \frac{K^2 \rho_0^2}{4\Gamma_0} \qty(\frac{\Gamma_0}{\Gamma})^{3/2}. \label{eq:zetaEqNoRR}
\end{align}
The longitudinal motion of the electron has three terms corresponding to the wake phase velocity, the electron energy, and the electron betatron oscillations, respectively.
The latter two terms quickly decrease with acceleration, and eventually we reach the stage when the electron has such a high energy, that its phase slippage is determined only be the wake phase velocity, $\dv*{\xi}{t} = 1/(2\gamma_\phase^2)$.
However, at the initial stage, the sign of $\dv*{\xi}{t}$ can be both positive and negative.
The threshold value of $\Gamma_{\threshold}$ which separates these cases can be found from the condition $\dv*{\xi}{t} = 0$ at $t = 0$,
\begin{align}
    &\frac{1}{2\gamma_\phase^2} - \frac{1}{2\Gamma_\threshold^2} - \frac{K^2 \rho_0^2}{4\Gamma_\threshold} = 0,\\
    &\Gamma_\threshold = \frac{\gamma_\phase^2}{4} \left[K^2 \rho_0^2 + \sqrt{K^4 \rho_0^4 + \frac{16}{\gamma_\phase^2}} \right].
\end{align}
There are two limiting cases.
When the initial betatron oscillations are weak enough, $K^2 \rho_0^2 \gamma_\phase \ll 1$, then we get a straightforward condition $\Gamma_\threshold = \gamma_\phase$.
If betatron oscillations are strong, $K^2 \rho_0^2 \gamma_\phase \gg 1$, then
\begin{equation}
    \Gamma_\threshold = \frac{1}{2} K^2 \rho_0^2 \gamma_\phase^2
\end{equation}
scales as $\gamma_\phase^2$.
Depending on whether the initial electron energy $\Gamma_0$ is larger or smaller then $\Gamma_\threshold$, the description will be different.

Let us assume that $\Gamma_0 > \Gamma_\threshold$ is satisfied.
In this case, $\dv*{\xi}{t}$ is positive from the very beginning.
As $\Gamma$ grows due to acceleration, the electron quickly reaches the regime when $\dv*{\xi}{t} \approx 1/(2\gamma_\phase^2)$.
If we neglect the transition to this regime, then we get the solution
\begin{align}
    &\xi = \xi_0 + \frac{t}{2\gamma_\phase^2},\\
    &\Gamma(\xi) = \Gamma_0 + 2\gamma_\phase^2 \int_{\xi_0}^\xi f(\xi) \dd{\xi}.
    \label{eq:gammaInitialFast}
\end{align}
So, a ``fast'' electron gets the same increase in energy regardless of its initial energy.

When $\Gamma_0 < \Gamma_\threshold$ (a ``slow'' electron), we have to do calculations for the part when the electron lags behind the wave ($\xi < \xi_0$).
We simplify the calculations assuming that the accelerating force is linear around $\xi_0$: $f(\xi) \approx f_0 - (\xi - \xi_0) / l$.
In this case, Eqs.~(\ref{eq:gammaEqNoRR}--\ref{eq:zetaEqNoRR}) reduce to the nonlinear oscillator equation,
\begin{equation}
    \dv[2]{\Gamma}{t} + \frac{1}{l} \left[\frac{1}{2\gamma_\phase^2} - \frac{1}{2\Gamma^2} - \frac{K^2 \rho_0^2}{4\Gamma_0} \qty(\frac{\Gamma_0}{\Gamma})^{3/2} \right].
\end{equation}
This equation has the first integral,
\begin{multline}
    \frac{1}{2} \qty(\dv{\Gamma}{t})^2 + \frac{1}{2l} \left[\frac{\Gamma}{\gamma_\phase^2} + \frac{1}{\Gamma} + K^2 \rho_0^2 \sqrt{\frac{\Gamma_0}{\Gamma}} \right] 
    \\= \frac{f_0^2}{2} + \frac{1}{2l} \left[\frac{\Gamma_0}{\gamma_\phase^2} + \frac{1}{\Gamma_0} + K^2 \rho_0^2 \right],
\end{multline}
The accelerated electron eventually gets sufficient longitudinal velocity to overtake the wave and return to the starting point $\xi = \xi_0$ with the energy $\tilde{\Gamma}_0$.
At this point $\dv*{\Gamma}{t} = f_0$, so the conservation of the integral gives us
\begin{multline}
    \frac{\tilde{\Gamma}_0}{\gamma_\phase^2} + \frac{1}{\tilde{\Gamma}_0} + K^2 \rho_0^2 \sqrt{\frac{\Gamma_0}{\tilde{\Gamma}_0}} \\   = \frac{\Gamma_0}{\gamma_\phase^2} + \frac{1}{\Gamma_0} + K^2 \rho_0^2.
\end{multline}
This equation (after multiplication by $\tilde\Gamma_0$) is a fourth-order algebraic equation with respect to $\tilde{\Gamma}_0^{1/2}$ and has one known root equal to $\Gamma_0^{1/2}$, so it can be reduced to a third-order equation
\begin{equation}
    \tilde{\Gamma}_0 = \gamma_\phase^2 \left[ \frac{1}{\Gamma_0} + K^2 \rho_0^2 \frac{\sqrt{\tilde{\Gamma}_0}}{\sqrt{\tilde{\Gamma}_0} + \sqrt{\Gamma_0}}\right].
\end{equation}
While it is possible to find exact solutions using the Cardano's method, we can find an approximate solution by employing the iterative procedure,
\begin{align}
    &\tilde{\Gamma}_{0,n} = \gamma_\phase^2 \left[ \frac{1}{\Gamma_0} + K^2 \rho_0^2 \alpha_{n-1}\right], \\
    &\alpha_n = \frac{\sqrt{\tilde{\Gamma}_{0,n}}}{\sqrt{\tilde{\Gamma}_{0,n}} + \sqrt{\Gamma_0}}, \quad \alpha_0 = 1.
\end{align}
We see that $\alpha_n$ is always between $0$ and $1$ by definition, and it is reduced with every iteration, so this procedure converges.
As usually $\tilde{\Gamma}_0 \gg \Gamma_0$, it is often sufficient to use the first $\tilde{\Gamma}_{0,1}$ or the second iteration $\tilde{\Gamma}_{0,2}$ as a sufficiently good approximation to the solution of $\tilde{\Gamma}_0$.

Let us find some limiting cases.
If betatron oscillations are negligible, $K^2 \rho_0^2 \Gamma_0 \ll 1$, then
\begin{equation}
    \tilde{\Gamma}_0 \approx \frac{\gamma_\phase^2}{\Gamma_0}.
\end{equation}
If betatron oscillations are strong enough, $K^2 \rho_0^2 \Gamma_0 \gg 1$, and $\tilde{\Gamma}_0 \gg \Gamma_0$, then
\begin{equation}
    \tilde{\Gamma}_0 \approx \frac{1}{2} K^2 \rho_0^2 \gamma_\phase^2 = \Gamma_\threshold.
\end{equation}

As soon as we have found $\tilde{\Gamma}_0$, we can apply formula \eqref{eq:gammaInitialFast} for a ``fast'' electron to $\tilde{\Gamma}_0$ as the new initial parameter and get
\begin{multline}
    \Gamma(\xi) = \tilde{\Gamma}_0 + 2\gamma_\phase^2 \int_{\xi_0}^\xi f(\xi) \dd{\xi} \\
    = \gamma_\phase^2 \left[\frac{1}{\Gamma_0} + \alpha K^2 \rho_0^2 + 2 \int_{\xi_0}^\xi f(\xi') \dd{\xi'} \right].
    \label{eq:gammaInitialSlow}
\end{multline}
This value is strictly proportional to $\gamma_\phase^2$, as expected of acceleration in a plasma wakefield.

\begin{figure}[tb!]
    \includegraphics[width=\linewidth]{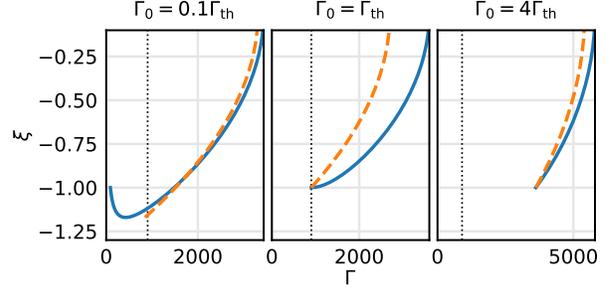}
    \caption{The numerical dependencies (solid) of $\Gamma$ on the coordinate $\xi$ for initial values $\Gamma_0 = 0.1 \Gamma_\threshold$, $\Gamma_\threshold$, $4\Gamma_\threshold$ calculated according to Eqs.~(\ref{eq:gammaEqNoRR}--\ref{eq:zetaEqNoRR}) and the corresponding analytical dependencies (dashed) based on either Eq.~\eqref{eq:gammaInitialSlow} or \eqref{eq:gammaInitialFast} with using the $\tilde{\Gamma}_{0,2}$ approximation.
    Parameters of the wake: $K^2 = 1/2$, $f(\xi) = - \xi/2$, $\gamma_\phase = 60$.
    Initial parameters of the electron are $\rho_0 = 1$, $\xi_0 = -1$, corresponding to $\Gamma_\threshold \approx 904$.
    The radiation reaction force is neglected.
    }
    \label{fig:initial_gamma}
\end{figure}

So, depending on whether the initial $\Gamma_0$ is lower or higher than the threshold value $\Gamma_\threshold$, we must use either Eq.~\eqref{eq:gammaInitialFast} with $\Gamma_0$ or Eq.~\eqref{eq:gammaInitialSlow} with $\tilde{\Gamma}_0$ as the initial value at $\xi_0$ and neglect the contribution from the electron parameters to phase slippage.
Figure~\ref{fig:initial_gamma} shows an example of the comparison of the numerical solution to analytical estimates for various values of $\Gamma_0$.
Both when $\Gamma_0 \ll \Gamma_\threshold$ or when $\Gamma_0 \gg \Gamma_\threshold$, the analytical estimates according to Eqs.~\eqref{eq:gammaInitialFast} and \eqref{eq:gammaInitialSlow} are close to the real solution.
In the worst case, when $\Gamma_0 = \Gamma_\threshold$, the analytical estimate undervalues $\Gamma$ due to overvaluing the rate of dephasing.

\section{Monte Carlo simulations}
\label{sec:monte-carlo}

To compare the classical description of radiation reaction used in the paper to probabilistic radiation reaction based on emission of individual photons, we have performed simulations using the Monte Carlo module from the software package QUILL \cite{Quill} used for particle-in-cell simulations.
In the simulations, we used analytical external EM fields $E_x = -B_y = x/4$, $E_y = B_x = y/4$, $E_z = (z - v_\phase t) / 2$, $B_z = 0$ which correspond to an ideally spherical model of a strongly nonlinear wakefield \cite{Kostyukov_2004_PoP_11_5256}.
Such fields provide the focussing constant $K^2 = 1/2$ and the accelerating force $f = \zeta / 2$.
The phase velocity corresponded to $\gamma_\phase = 400$, and the plasma density used in normalization was equal to \SI{e23}{\cm^{-3}}.
A cylindrical electron bunch with the length $\zeta_\bunch = 1.2\pi$, radius $r_\bunch = 1.6\pi$, and initial $\gamma_0 = 100$ (with initial momentum along the $z$-axis) was placed inside the accelerating phase of the wakefield.
Current deposition and generation of EM fields by electrons were turned off, so the electrons did not interact with each other.

\begin{figure}[tb]
    \includegraphics[width=\linewidth]{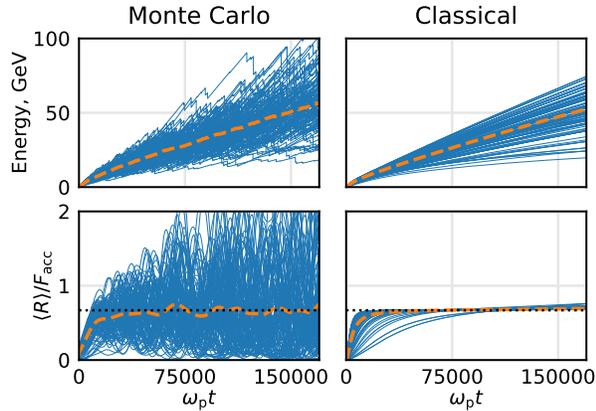}
    \caption{
        Dependencies of the energies of accelerated electrons and the relative radiation reaction force in a bunch in Monte Carlo simulations and in the solution using the classical radiation reaction force.
        The dashed lines show the averaged over the bunch values of the energy and relative radiation reaction force.
    }
    \label{fig:pic}
\end{figure}

The comparison between the Monte Carlo simulations and the solution with the same initial conditions for the radiation reaction force in the classical limit for a subset of the bunch particles is shown in Fig.~\ref{fig:pic}.
In Monte Carlo simulations, each particle experiences abrupt drops in energy as it radiates individual photons, and the visible energy spread is slightly larger in Monte Carlo simulations due to stochasticity.
However, the energy averaged over all bunch particles has the same dependence in both cases.
The radiation reaction force in the Monte Carlo simulations is poorly defined, as abrupt changes in energy correspond to a series of $\delta$-functions (or $\delta$-like pulses for the discretized solution).
Therefore, to visually represent the average radiation reaction force, we used convolution of these spikes with the Blackman window function of large width $8000\pi \approx 25000$.
Even then, such an averaged force rapidly oscillates, demonstrating the stochastic behavior of radiation reaction.
However, the average value of the force over all particles is very close to the value $2/3$ predicted by the radiation reaction force in the classical limit.
    
\bibliographystyle{aipnum4-1}
\bibliography{Bibliography}

\end{document}